\documentclass[sigconf,authorversion,nonacm]{acmart}

\usepackage{comment}
\usepackage{graphicx}
\usepackage{caption}
\usepackage{subcaption}
\usepackage{url}
\usepackage{xurl}
\usepackage{hyperref}
\usepackage{listings}
\usepackage{float}

\AtBeginDocument{%
  \providecommand\BibTeX{{%
    \normalfont B\kern-0.5em{\scshape i\kern-0.25em b}\kern-0.8em\TeX}}}

\begin{document}

\title{Extracting Information from Twitter Screenshots}

\author{Tarannum Zaki}
\affiliation{%
  \institution{Old Dominion University}
  \city{Norfolk}
  \state{Virginia}
  \country{USA}
}
\email{tzaki001@odu.edu}

\author{Michael L. Nelson}
\affiliation{%
  \institution{Old Dominion University}
  \city{Norfolk}
  \state{Virginia}
  \country{USA}
}
\email{mln@cs.odu.edu}
\author{Michele C. Weigle}
\affiliation{%
  \institution{Old Dominion University}
  \city{Norfolk}
  \state{Virginia}
  \country{USA}
}
\email{mweigle@cs.odu.edu}

\renewcommand{\shortauthors}{Zaki et al.}

\begin{abstract}
Screenshots are prevalent on social media as a common approach for information sharing. Users rarely verify before sharing a screenshot whether the post it contains is fake or real. Information sharing through fake screenshots can be highly responsible for misinformation and disinformation spread on social media. Our ultimate goal is to develop a tool that could take a screenshot of a tweet and provide a probability that the tweet is real, using resources found on the live web and in web archives. This paper provides methods for extracting the tweet text, timestamp, and Twitter handle from a screenshot of a tweet.
\end{abstract}

%\begin{CCSXML}
%<ccs2012>
%  <concept_id>10010520.10010553.10010562</concept_id>
%  <concept_desc>Computer systems organization~Embedded systems</concept_desc>
%  <concept_significance>500</concept_significance>
%  <concept_id>10010520.10010575.10010755</concept_id>
%  <concept_desc>Computer systems organization~Redundancy</concept_desc>
%  <concept_significance>300</concept_significance>
 % <concept_id>10010520.10010553.10010554</concept_id>
 % <concept_desc>Computer systems organization~Robotics</concept_desc>
 % <concept_significance>100</concept_significance>
 %</concept>
 %<concept>
 % <concept_id>10003033.10003083.10003095</concept_id>
 % <concept_desc>Networks~Network reliability</concept_desc>
 % <concept_significance>100</concept_significance>
 %</concept>
%</ccs2012>
%\end{CCSXML}

%\ccsdesc[500]{Computer systems organization~Embedded systems}




\maketitle

\section{Introduction}
\begin{figure}
    \includegraphics[scale=0.43]{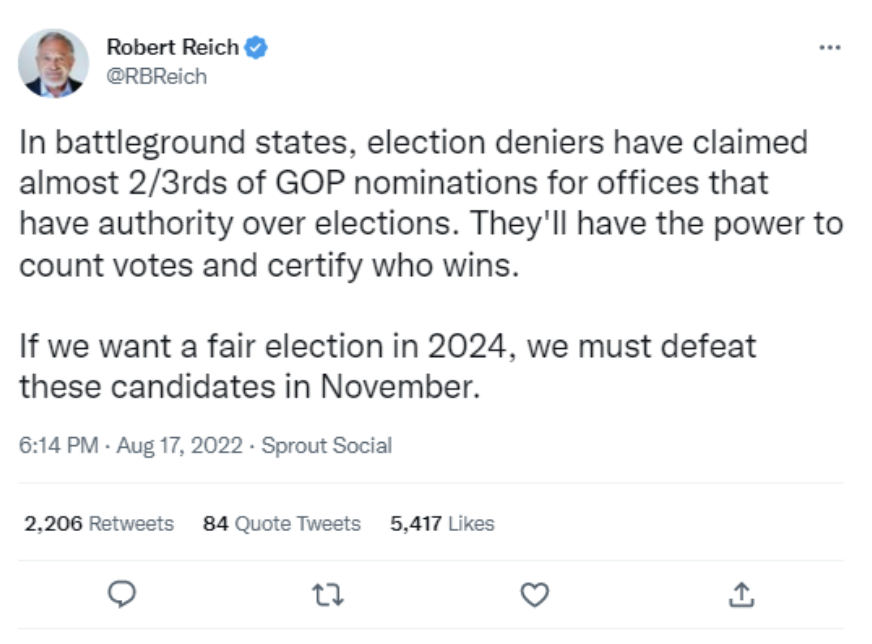}
    \caption{Tweet posted by @RBReich (\url{https://twitter.com/RBReich/status/1560027191404072961}).}
    \label{reich-tweet}
\end{figure}

\begin{figure}
    \includegraphics[scale=0.43]{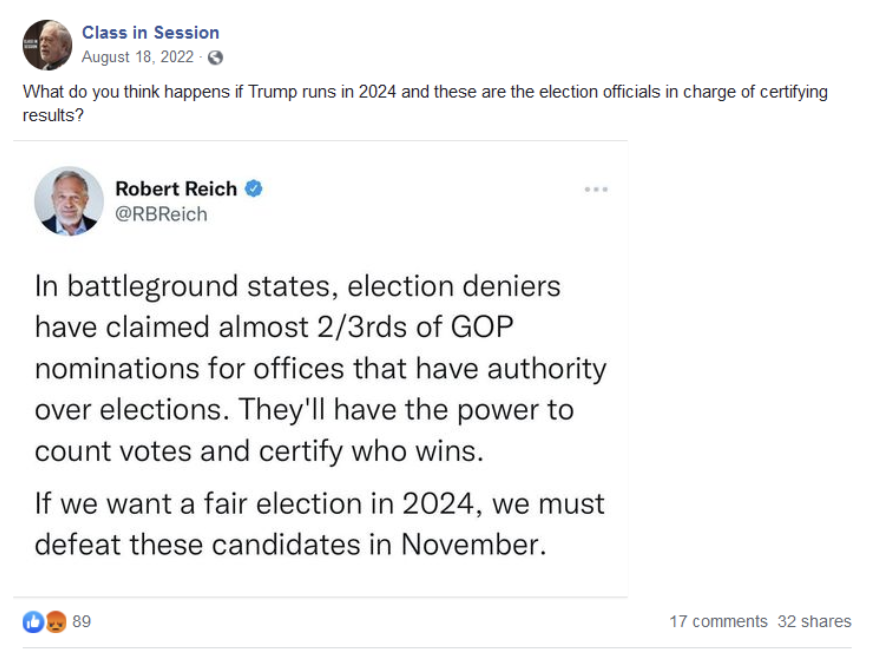}
    \caption{Screenshot of tweet posted on Facebook by @RBReich (\url{https://www.facebook.com/watchclassinsession/posts/pfbid0344Hu2bxJtAiiL5VHfM2YQyPTU9jTm3tfdJMj4TZMDunomMarXMQfTxPGvsVwfBmwl}).}
    \label{reich-tweet-facebook}
\end{figure}

\begin{figure}[htp]
\centering
  \begin{subfigure}{0.30\textwidth}
  \centering
   \includegraphics[width=\linewidth]{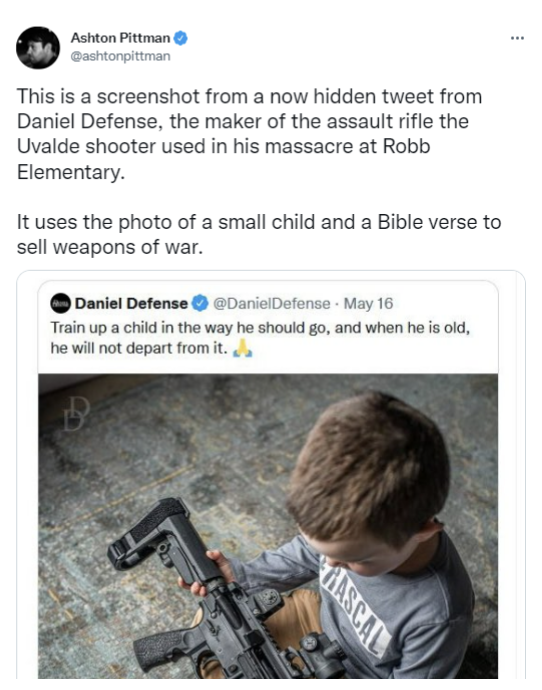}
    \caption{Screenshot of deleted tweet from @DanielDefense posted by @ashtonpittman (\url{https://twitter.com/ashtonpittman/status/1530243294868930560}).}
    \label{deleted-tweet-defense}
  \end{subfigure}
  %\hfill %%
  \begin{subfigure}{0.30\textwidth}
  \centering
    \includegraphics[width=\linewidth]{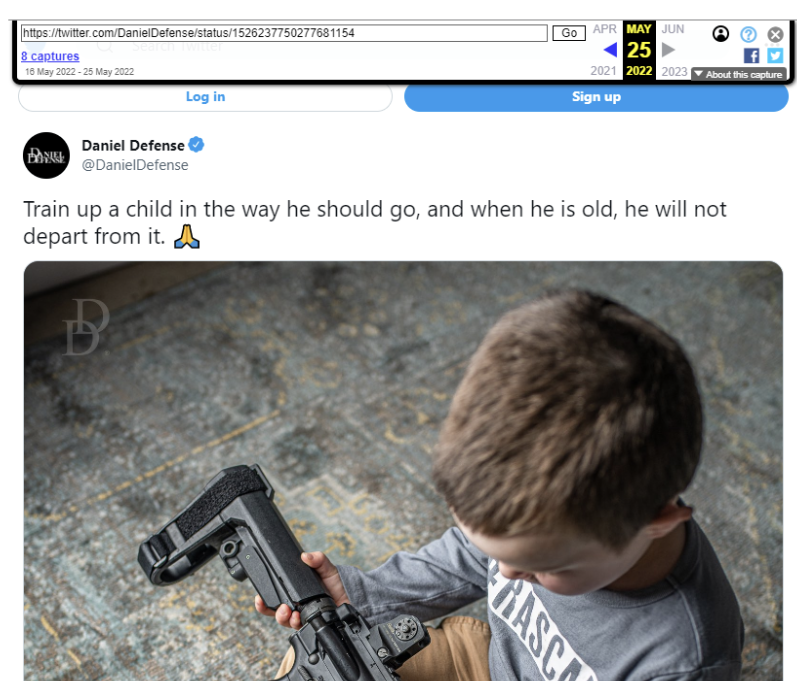}
    \caption{Archived version of the deleted tweet by @DanielDefense (\url{https://web.archive.org/web/20220525125749/https://twitter.com/DanielDefense/status/1526237750277681154}).}
    \label{archive-tweet-defense}
  \end{subfigure}
\caption{Screenshot of a deleted tweet and it's archived version.}
\end{figure}

A screenshot is a way to share content on social media that allows cross-platform user engagement. For example, @RBReich shared his own tweet (Figure \ref{reich-tweet}) as a screenshot on Facebook (Figure \ref{reich-tweet-facebook}) to increase cross-platform engagement \cite{reich-facebook-2022}. Moreover, screenshots are also used as evidence where there are chances of controversial posts being deleted. For example, a tweet posted by @DanielDefense about the Uvalde shooting incident was deleted but a screenshot of the deleted tweet was shared as evidence (Figure \ref{deleted-tweet-defense}). 

\begin{comment}
\begin{figure}
    \includegraphics[scale=0.5]{images/Fig3.png}
    \caption{Screenshot of deleted tweet from @DanielDefense posted by @ashtonpittman (\url{https://twitter.com/ashtonpittman/status/1530243294868930560}).}
    \label{deleted-tweet-defense}
\end{figure}
\end{comment}

\begin{figure}[htp]
\centering
  \begin{subfigure}[b]{0.35\textwidth}
  \centering
    \includegraphics[width=\linewidth]{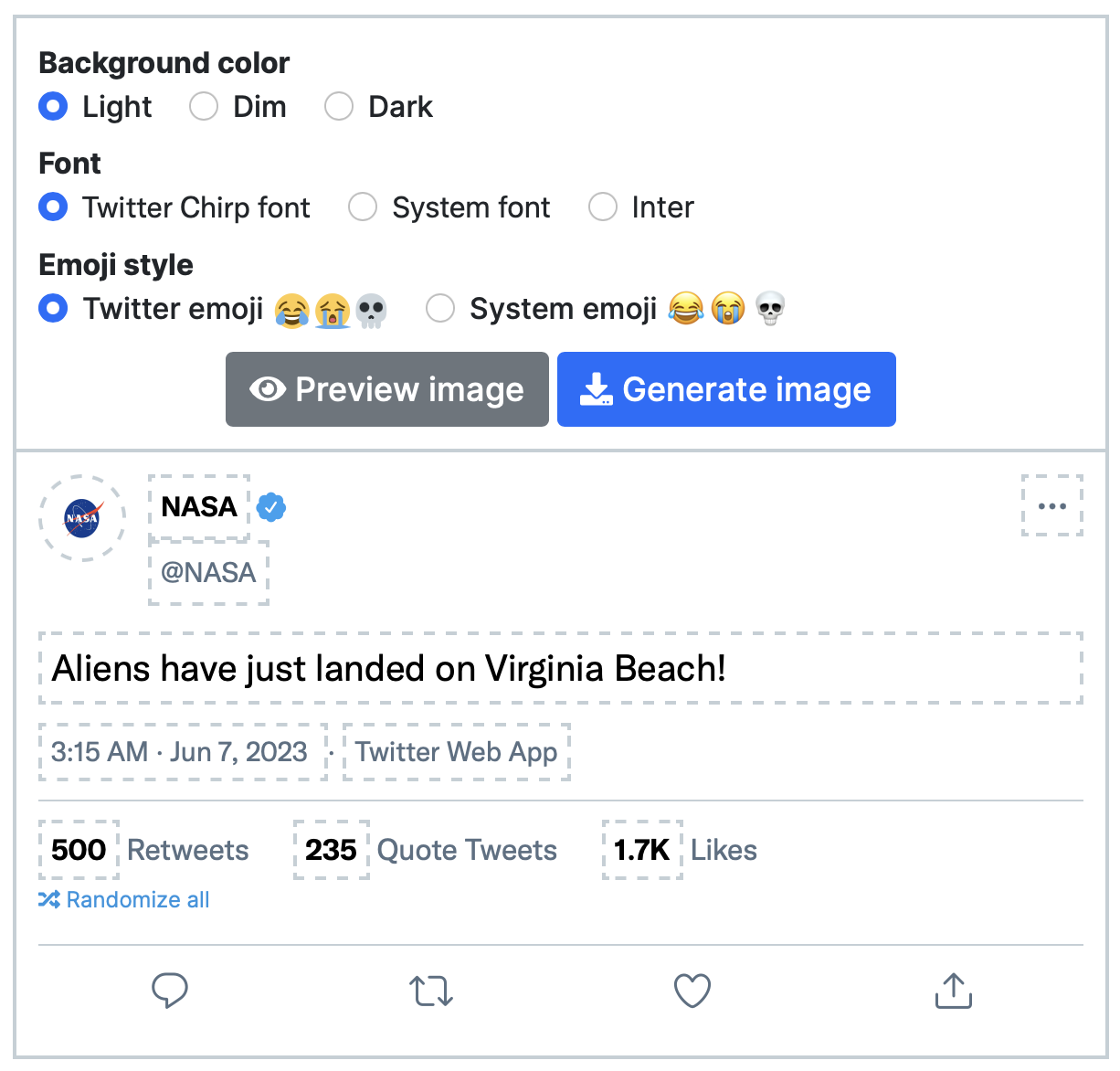}
    \caption{Editable attributes to create a fake tweet from @NASA.}
    \label{create-fake-tweet}
  \end{subfigure}
  \vspace{4ex} %%
  \begin{subfigure}[b]{0.35\textwidth}
  \centering
    \includegraphics[width=\linewidth]{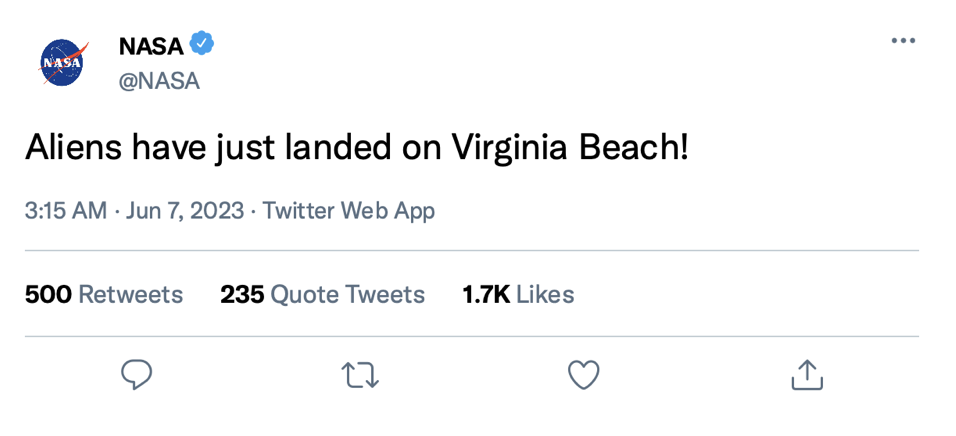}
    \caption{A fake tweet from @NASA.}
    \label{fake-tweet}
  \end{subfigure}
\caption{An example fake tweet created by using Tweetgen.}
\end{figure}

The wide use of screenshots is largely because it is easy to take a screenshot. Moreover, screenshots can be suspect because fake tweets are relatively easy to create using online tools such as Tweetgen.\footnote{https://www.tweetgen.com/} It is difficult to edit or delete a screenshot of a fake tweet once it has been shared across social media. So, it becomes quite challenging to detect whether the content of the screenshot is real or fake. Figures \ref{create-fake-tweet} and \ref{fake-tweet} show an example of how Tweetgen can be used to create a fake tweet. There are ways to edit metadata (e.g. handle name, display name, timestamp, profile picture, tweet, count of likes, retweets etc.) in this tool which results in creating an image that looks like an actual tweet.

There are no tools currently available to evaluate the authenticity of screenshots shared on social media, but there are methods which could be utilized to verify whether the content of a screenshot had been really posted by the author. The simplest way of searching on the live web is by searching for the text in a search engine, such as Google. Furthermore, there exist fact-checking websites, like Snopes\footnote{https://www.snopes.com/} and FactCheck.org,\footnote{https://www.factcheck.org/} where users can search whether content is fabricated. For example, a fabricated tweet by Rep. Marjorie Taylor Greene regarding the 4th of July was fact-checked by FactCheck.org \cite{spencer-factcheck-2022}. Another useful method in this regard is searching web archives such as the Wayback Machine for deleted posts or accounts. The previously mentioned account @DanielDefense is an example of finding deleted tweets on web archives. Figure \ref{archive-tweet-defense} shows the archived version of the deleted tweet \cite{defense-tweet-2022}.

\begin{comment}
\begin{figure}[ht]
\begin{subfigure}{.5\textwidth}
  \centering
  % include first image
  \includegraphics[width=.8\linewidth]{images/Fig3.png}  
  \caption{Screenshot of deleted tweet from @DanielDefense posted by @ashtonpittman (\url{https://twitter.com/ashtonpittman/status/1530243294868930560}).}
   \label{deleted-tweet-defense}
\end{subfigure}
\begin{subfigure}{.5\textwidth}
  \centering
  % include second image
  \includegraphics[width=.8\linewidth]{images/Fig7.png}  
  \caption{Archived version of the deleted tweet by @DanielDefense (\url{https://web.archive.org/web/20220525125749/https://twitter.com/DanielDefense/status/1526237750277681154}).} \label{archive-tweet-defense}
\end{subfigure}
\caption{An example fake tweet created by using Tweetgen.}
\label{fig:fig}
\end{figure}
\end{comment}

\begin{comment}
\begin{figure}
    \includegraphics[scale=0.5]{images/Fig7.png}
    \caption{Archived version of the deleted tweet by @DanielDefense (\url{https://web.archive.org/web/20220525125749/https://twitter.com/DanielDefense/status/1526237750277681154}).}
    \label{archive-tweet-defense}
\end{figure}
\end{comment}

People can browse on the live web using search engines and fact-checking websites to verify content of a screenshot. Similarly, they can look into the web archives. These are some manual ways that people can utilize the live web and web archives to validate whether the content of a screenshot is real or fake. Our goal is to automate this manual process to see if automation makes it easier to validate the content of a screenshot. We are developing an automated tool using such services to estimate the likelihood that the content of a shared screenshot on social media had been really posted by the alleged author. Our research will aid in alleviating misinformation and disinformation spread by detecting whether the content of a screenshot is fake or real. 

As a part of this overall goal, first we need to be able to extract certain pieces of information from a screenshot image of a tweet. Here, we provide methods for extracting tweet text, timestamp, and Twitter handle from a screenshot of a tweet and discuss the evaluation of the methods.

\section{Methodology}

\begin{figure*}[htp]
    \includegraphics[width=0.7\textwidth]{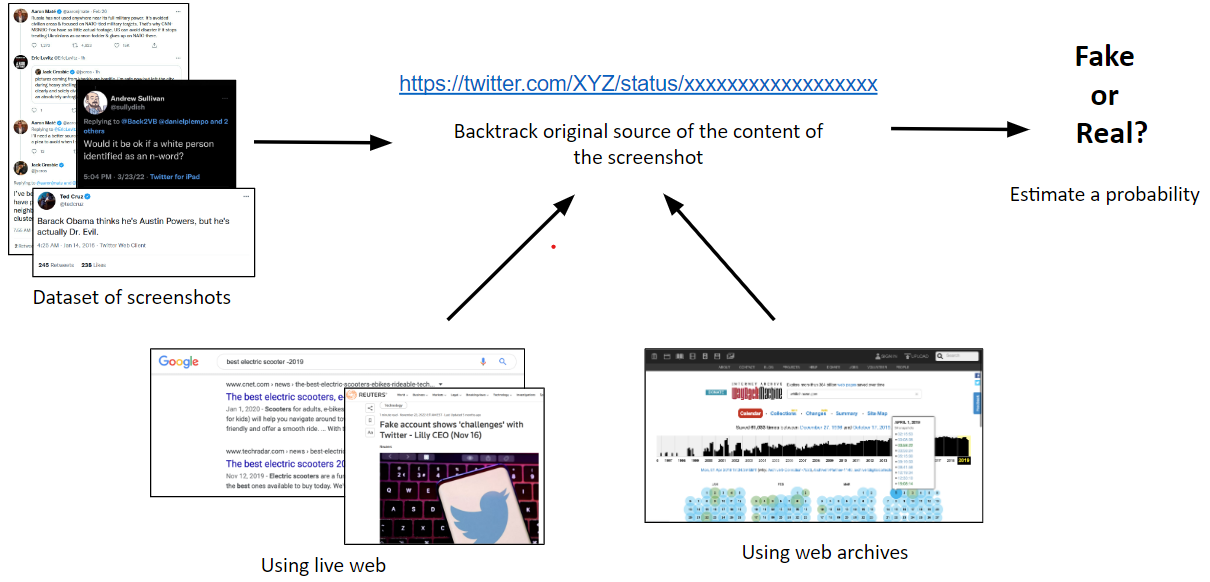}
    \caption{Methodology to predict whether a screenshot is fake or real.}
    \label{methodology}
\end{figure*}

Figure \ref{methodology} shows the methodology for the entire process to predict whether a screenshot is fake or real. The process starts with collecting screenshot images. Next, the search strategies are defined - searching content of a screenshot on the live web and web archives. Lastly, if the original source of the content is found, then the content could be verified as real. Otherwise, a probability would be estimated to verify whether the content is fake. 

We created a data set of screenshots posted on Twitter. The data set currently consists of 200 screenshot images with both real and fake examples. A detailed description of the data set is provided in our blog post \cite{zaki-blog-2022}.

\begin{figure}
    \includegraphics[scale=0.297]{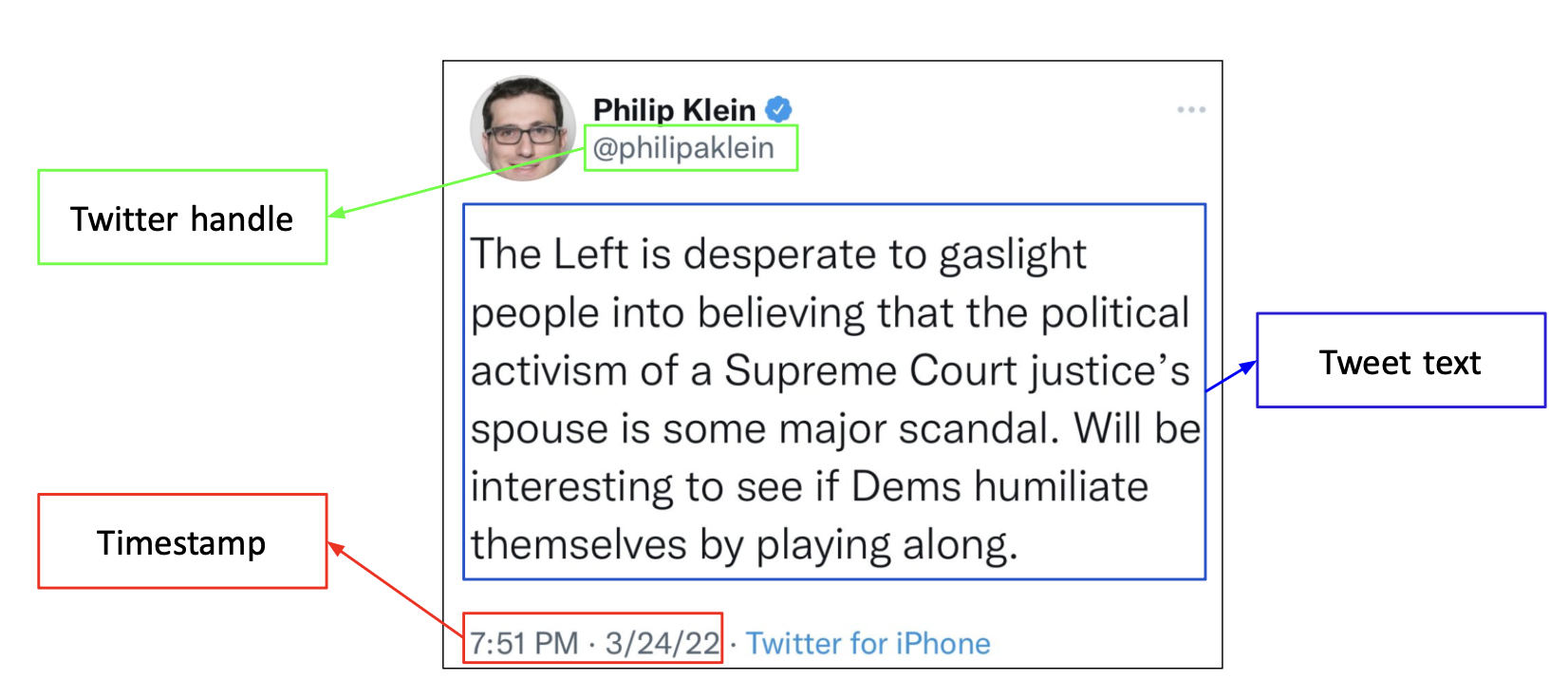}
    \caption{Information required to be extracted from a screenshot (\url{https://twitter.com/philipaklein/status/1507188518459777024}).}
    \label{tweet-info-image}
\end{figure}

The goal is to backtrack to the original link of the content of the screenshot. One way this can be done is by using the tweet text as a query to search engines and fact-checking websites on the live web. This requires the tweet text, Twitter handle, and timestamp be extracted to validate the screenshot (Figure \ref{tweet-info-image}).

\begin{figure*}[htp]
    \includegraphics[scale=0.37]{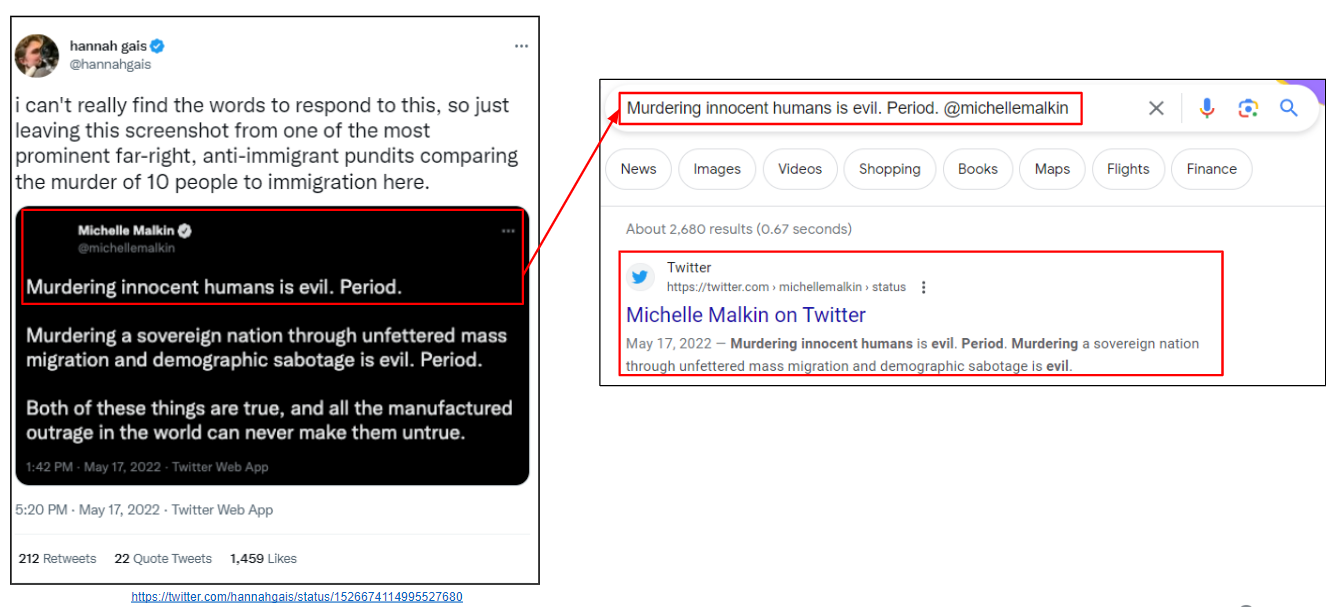}
    \caption{Using tweet text and Twitter handle to backtrack to the original URL using Google Search (\url{https://twitter.com/hannahgais/status/1526674114995527680}).}
    \label{google-search-image}
\end{figure*}

Figure \ref{google-search-image} shows how tweet text from a screenshot can be used to backtrack to the original URL using Google Search. Here, a substring (`Murdering innocent humans is evil. Period.') from the shared screenshot along with the Twitter handle (@michellemalkin) is used as a Google Search query to find whether anything relevant exists in the live web. The search returns the original Twitter URL of the post shared by @michellemalkin. This indicates that the content of the shared screenshot is real.

\begin{figure*}[htp]
    \includegraphics[scale=0.37]{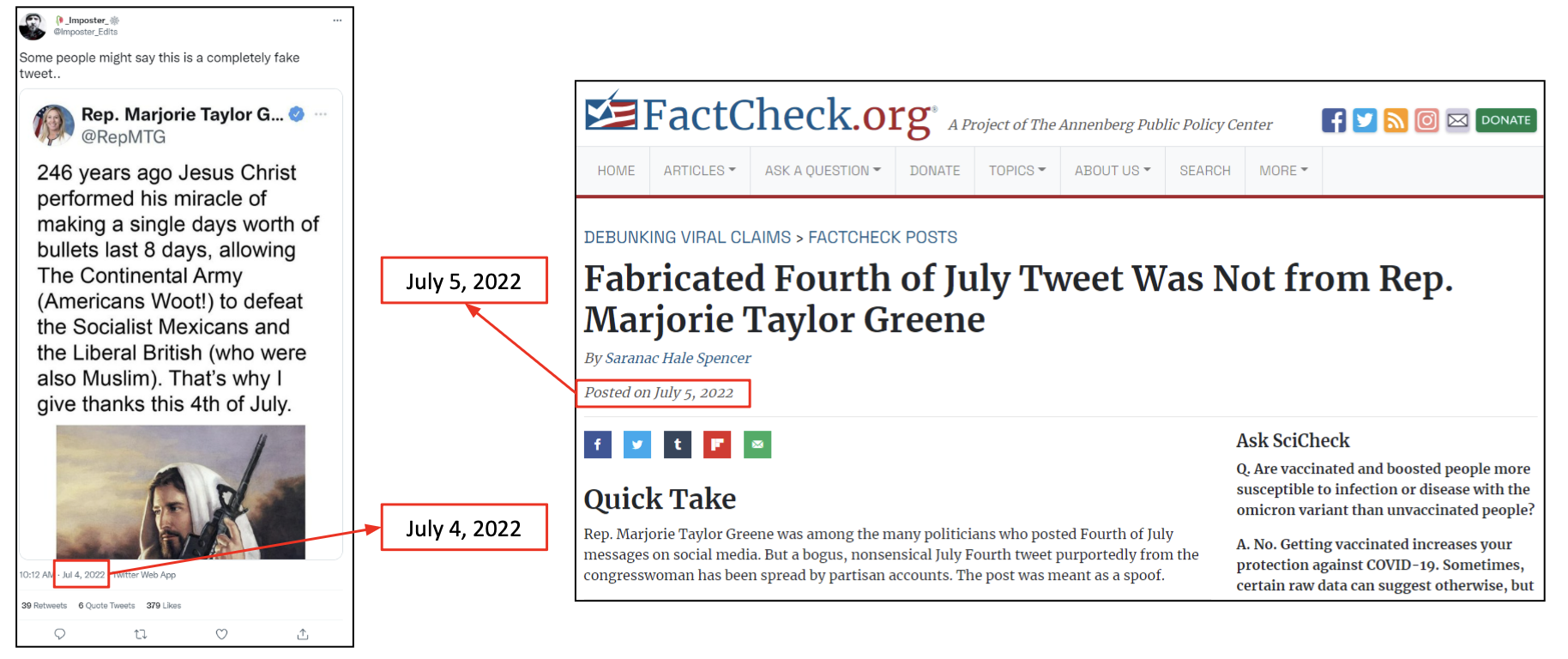}
    \caption{Using timestamp and tweet text to backtrack original URL using a fact-checking website (\url{https://twitter.com/Imposter_Edits/status/1543960895965085696}).}
    \label{factcheck-image}
\end{figure*}

Figure \ref{factcheck-image} shows how the timestamp and tweet text from a screenshot can be used to backtrack to the original URL using a fact-checking website. An example discussed earlier involves a screenshot of a fabricated post of @RepMTG that was shared on Twitter on July 4, 2002. The fact-checking website FactCheck.org verified on July 5, 2022 that the content was not posted by the alleged author. They checked the author’s official Twitter feed as well as on Politwoops\footnote{https://projects.propublica.org/politwoops/} - a database of politicians' deleted tweets. The verification process of FactCheck.org indicates that the content of the shared screenshot is fake.

Another way is to search web archives, such as the Wayback Machine. Most web archives are indexed by URLs. A Twitter URL is structured as `https://twitter.com/\emph{Twitter handle}/status/\emph{tweet ID}'. So, we require the Twitter handle from a screenshot image. Given the user's handle, we can find all of the archived tweets for that user from the Internet Archive using their CDX API\footnote{https://github.com/internetarchive/wayback/tree/master/wayback-cdx-server} with a query for `https://twitter.com/Twitter handle/status/*'. But, in order retrieve the exact URL, we need a likely time period, which we can obtain from the timestamp of a screenshot.

\begin{figure}[htp]
\centering
  \begin{subfigure}[b]{0.3\textwidth}
  \centering
    \includegraphics[width=\linewidth]{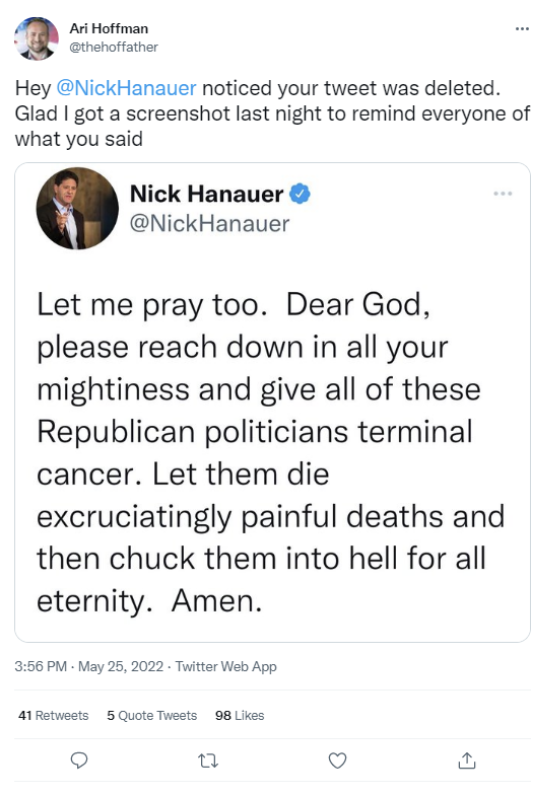}
    \caption{A deleted tweet posted by @NickHanauer.}
    \label{deleted-tweet-hanauer}
  \end{subfigure}
  \vspace{4ex} %%
  \begin{subfigure}[b]{0.43\textwidth}
  \centering
    \includegraphics[width=\linewidth]{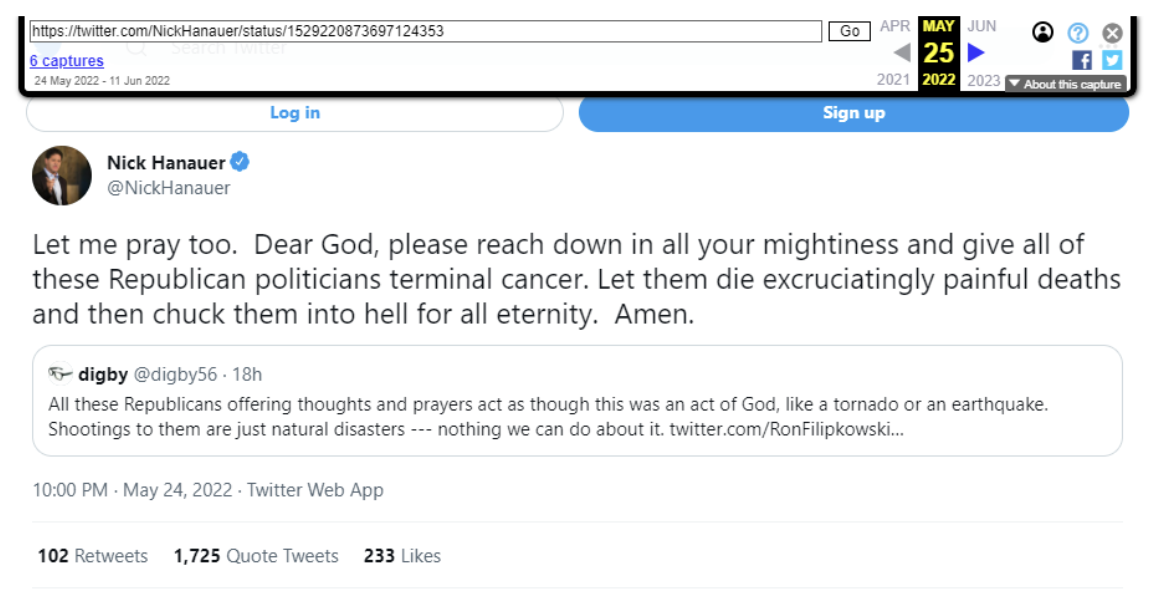}
    \caption{Archived version of a deleted tweet posted by @NickHanauer (\url{https://web.archive.org/web/20220525164026/https://twitter.com/NickHanauer/status/1529220873697124353}).}
    \label{archived-tweet-hanauer}
  \end{subfigure}
\caption{Screenshot of a deleted tweet and it's archived version.}
\end{figure}

\begin{comment}
\begin{figure}
    \includegraphics[scale=0.45]{images/Fig12.png}
    \caption{A deleted tweet posted by @NickHanauer.}
    \label{deleted-tweet-hanauer}
\end{figure}
\end{comment}

Figure \ref{deleted-tweet-hanauer} shows a screenshot of a tweet \cite{hanauer-tweet-2022}. The tweet was originally posted by @NickHanauer but later deleted from the Twitter feed. From this, we can get the Twitter handle (@Nickhanauer) and the timestamp (May 25, 2022). This indicates that if the tweet had been archived, then a possible timestamp range would be from May 25, 2022 to May 26, 2022. We use the curl command, as shown in \ref{code-snippet} to access the CDX API, where the URL parameter is set to the URL (https://twitter.com/NickHanauer/status) we want to search in the web archive. The timestamp field is set to the corresponding timestamp range (from=20220525, to=20220526) in order to retrieve any relevant archived URLs within that time frame. The archived version of the URL for the deleted tweet is colored in red in the list of retrieved URLs. Figure \ref{archived-tweet-hanauer} shows the retrieved URL, which is the archived version of the deleted tweet by @NickHanauer.

\begin{figure}
\lstset{moredelim=[is][\color{red}]{[*}{*]}}
\begin{lstlisting}[breaklines, basicstyle=\small]
curl -s "http://web.archive.org/cdx/search/cdx?url=https://twitter.com/NickHanauer/status&from=20220525&to=20220526&matchType=prefix" | sort -u -k 3 | awk '{print "https://web.archive.org/web/" $2 "/" $3};'

https://web.archive.org/web/20220525153810/https://twitter.com/NickHanauer/status/1305869227409027072
https://web.archive.org/web/20220526062353/https://twitter.com/NickHanauer/status/1305869227409027072
https://web.archive.org/web/20220526035516/https://twitter.com/NickHanauer/status/1305869227409027072
https://web.archive.org/web/20220525184648/https://twitter.com/NickHanauer/status/1305869227409027072
https://web.archive.org/web/20220525205256/https://twitter.com/NickHanauer/status/1374401501024583683
[*https://web.archive.org/web/20220525164026/https://twitter.com/NickHanauer/status/1529220873697124353*]
\end{lstlisting}
\caption{Code snippet to retrieve URLs from the Wayback Machine using the CDX API.}
\label{code-snippet}
\end{figure}

\begin{comment}
\begin{figure}
    \includegraphics[scale=0.4]{images/Fig13.png}
    \caption{Archived version of a deleted tweet posted by @NickHanauer (\url{https://web.archive.org/web/20220525164026/https://twitter.com/NickHanauer/status/1529220873697124353}).}
    \label{archived-tweet-hanauer}
\end{figure}
\end{comment}

If content had really been posted by the alleged author, there is the possibility of finding it on the live web and web archives. This would help to determine whether the screenshot is valid.

\section{Results and Discussion}

\begin{figure}[htp]
    \includegraphics[scale=0.25]{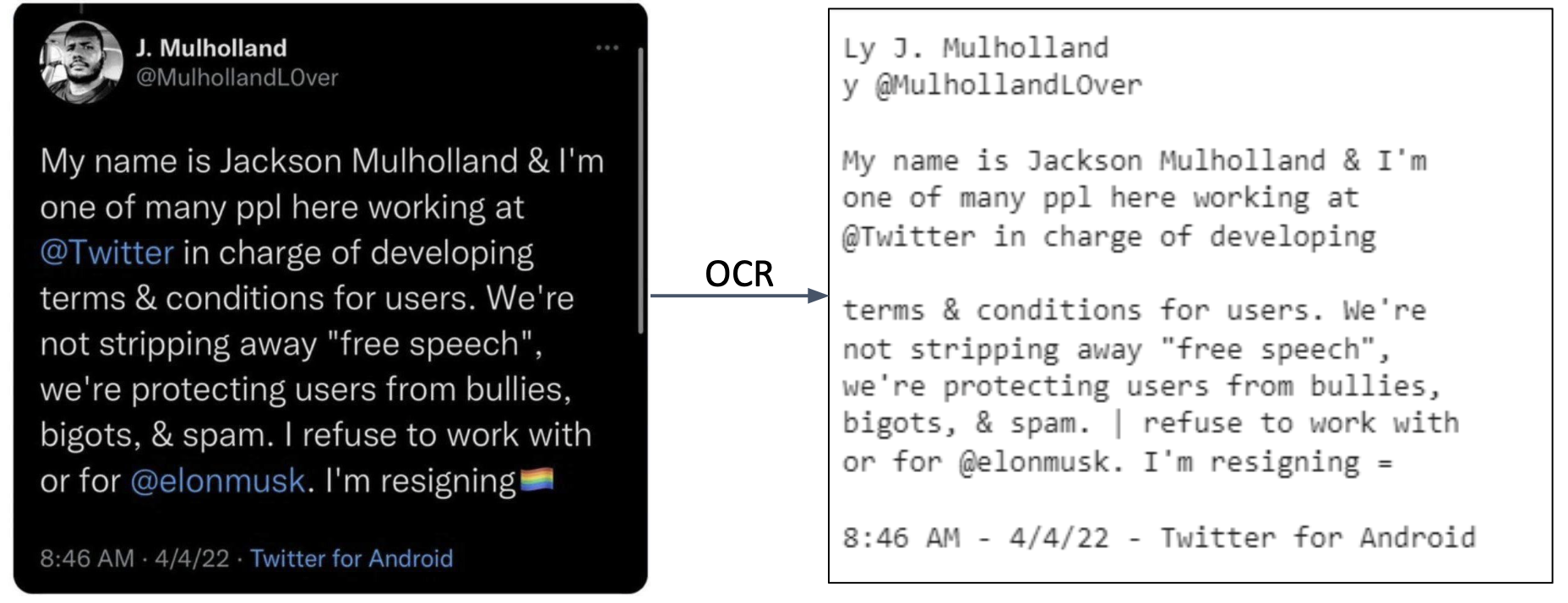}
    \caption{Applying OCR to convert text from a screenshot image.}
    \label{ocr-conversion}
\end{figure}

The input for the intended tool is a screenshot image. Initially, optical character recognition (OCR) is applied to extract the information available in the screenshot. The Python-Tesseract\footnote{https://pypi.org/project/pytesseract/} tool is used to perform the OCR. This tool recognizes any textual description that is embedded on the images and outputs the corresponding text. Figure \ref{ocr-conversion} shows how OCR converts text from a digital image of a screenshot. The entire tweet text along with the Twitter handle and timestamp are retrieved from this OCR converted output of a screenshot image. 

Our search strategies require the extraction of the timestamp, Twitter handle, and tweet text particularly. So far, we have performed an experiment for extracting this information from the OCR converted text using 125 single tweet screenshot images from our data set\cite{zaki-dataset}. By single tweet images we are referring to images that include only a single tweet (e.g. screenshot image of Figure \ref{ocr-conversion}) and do not involve any quote tweet, thread, or other concatenated part of tweets. 

\begin{figure*}[htp]
    \includegraphics[scale=0.43]{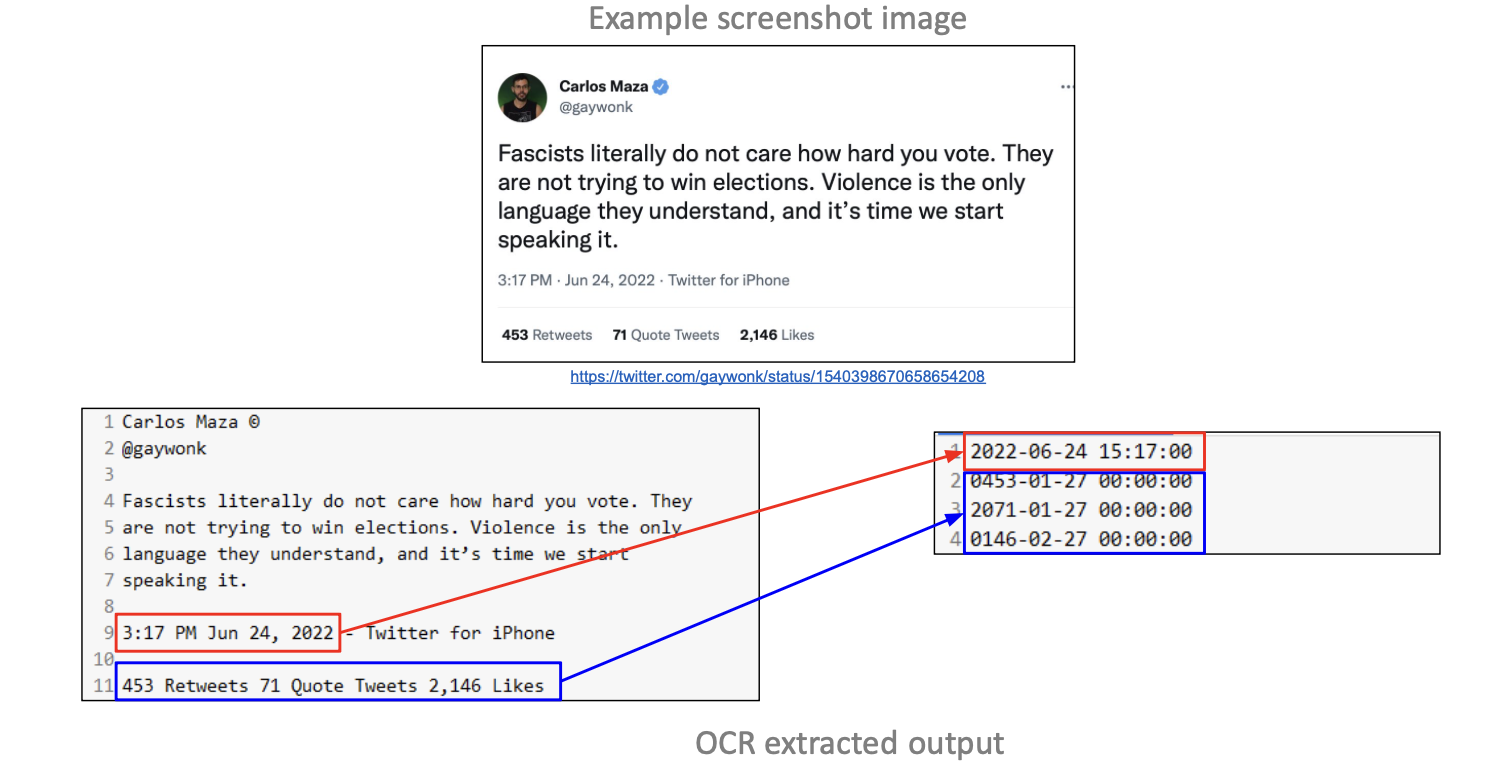}
    \caption{Using Method 1 to extract timestamp from screenshot image.}
    \label{method1-timestamp}
\end{figure*}

For extracting the timestamp of a post from a screenshot, two methods have been evaluated. Method 1 uses the Python module datefinder \footnote{https://pypi.org/project/datefinder/} for extracting the timestamp, whereas Method 2 uses a logic on the date format along with it. The datefinder module extracts date type strings from a text. Figure \ref{method1-timestamp} demonstrates Method 1, showing which shows the timestamp from the OCR converted output (3:17 PM Jun 24, 2022) is extracted as 2022-06-24 15:17:00 (annotated with red colored box). But, there are some discrepancies using this method. This method also extracts other numerical digits as random timestamps (annotated with blue colored box), which are incorrect (e.g. 0453 Retweets is extracted as 0453-01-27 00:00:00). 
In order to deal with the discrepancy while extracting the timestamp from the screenshot images, we use additional logic in the date format. We required at least 6 characters (including the separator) and at least 4 digits to define a fully described date for this logic. On the other hand, Method 2, extracts the timestamp from the OCR converted output (3:17 PM Jun 24, 2022) as 2022-06-24 15:17:00 only. As the text in blue colored box does not meet the criteria, so the text is red colored box is retrieved only.
\begin{comment}
\begin{figure}
    \includegraphics[scale=0.33]{images/Fig17.png}
    \caption{Using Method 2 to extract timestamp from screenshot image.}
    \label{method2-timestamp}
\end{figure}
\end{comment}
We evaluated the two methods on timestamp extraction, and Table \ref{table-timestamp} shows that Method 2 performs better than Method 1 in terms of accuracy, precision, recall, and F1 score.

\begin{table}[htbp]
  \caption{Performance evaluation of methods for extracting timestamp.}
  \label{table-timestamp}
  \begin{tabular}{ccccc}
    \toprule
    \textbf{Methods} & \textbf{Accuracy} & \textbf{Precision} & \textbf{Recall} & \textbf{F1 Score} \\
    \midrule
    \textbf{Method 1}&40\%&60\%&39\%&47\% \\
    \textbf{Method 2}&80\%&74\%&97\%&89\% \\
    \bottomrule
  \end{tabular}
\end{table}

\begin{figure}
    \includegraphics[scale=0.4]{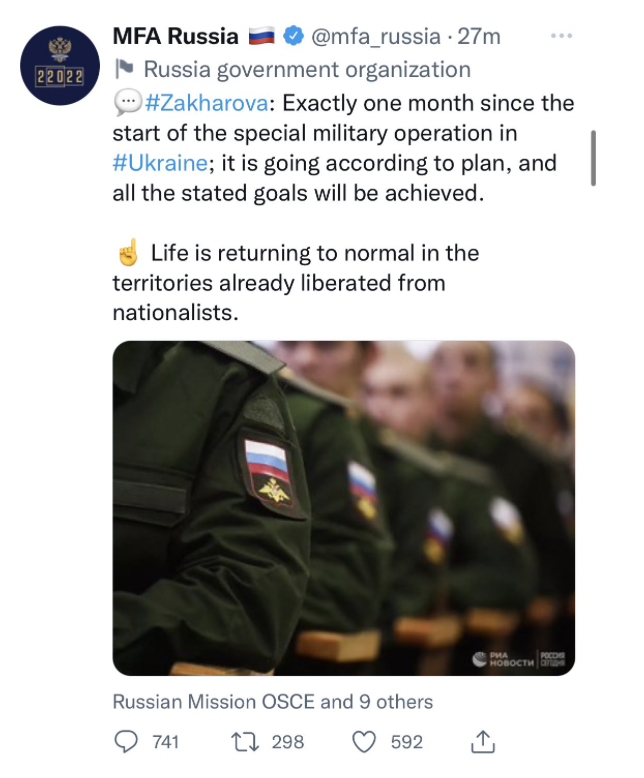}
    \caption{Example of a screenshot where it is not possible to extract the timestamp using either of our methods.}
    \label{timestamp-exception}
\end{figure}

\begin{figure}[htp]
    \includegraphics[scale=0.35]{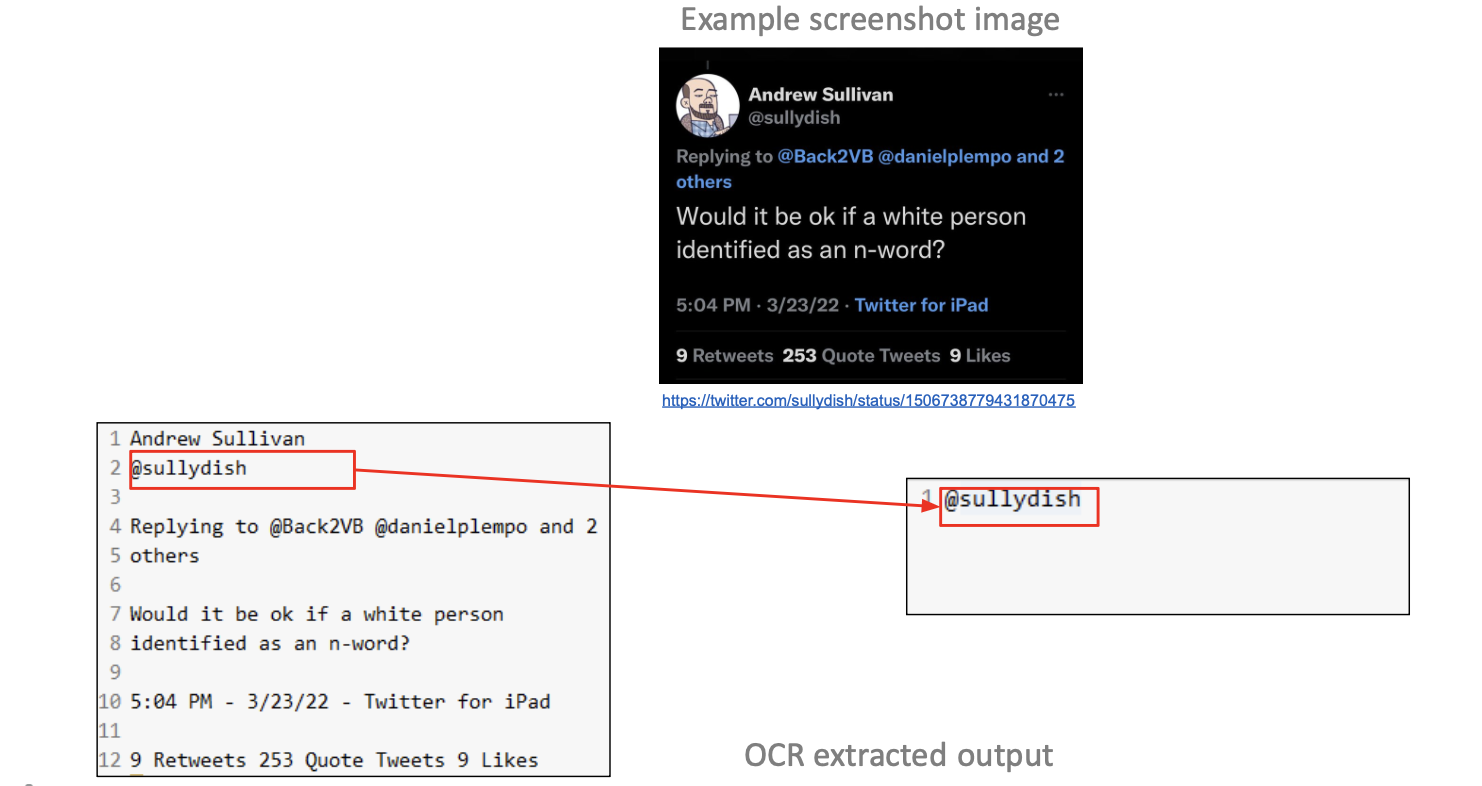}
    \caption{Extracting Twitter handle from screenshot image.}
    \label{twitter-handle}
\end{figure}

However, there are limitations to both of these methods. For example, screenshots that do not contain timestamp in a particular date time format will not result in any output. Figure \ref{timestamp-exception} shows such an example where the timestamp is specified as `27m' which which cannot be used to extract a full date time
For extracting the Twitter handle of a post from a screenshot, some aspects need to be considered. A Twitter handle starts with the symbol `@'. There might be multiple Twitter handles in a single tweet. For example, such a case is when other users are mentioned in a tweet. Moreover, the verified check mark is converted to a `@' symbol while using OCR. To handle these cases, we traverse through each line of the OCR extracted text, considering words that start with `@' that are not only `@’, and extracting the first one among such matched words. Figure \ref{twitter-handle} demonstrates this method and shows only the Twitter handle of the post is extracted from the OCR converted output.
We evaluated this method on Twitter handle extraction, and Table \ref{table-twitter-handle} shows the accuracy, precision, recall, and F1 score.

\begin{table}[htbp]
  \caption{Performance evaluation of the method for extracting the Twitter handle.}
  \label{table-twitter-handle}
  \begin{tabular}{cccc}
    \toprule
    \textbf{Accuracy} & \textbf{Precision} & \textbf{Recall} & \textbf{F1 Score} \\
    \midrule
     84\%&99\%&85\%&91\% \\
    \bottomrule
  \end{tabular}
\end{table}

\begin{figure}[htp]
    \includegraphics[scale=0.4]{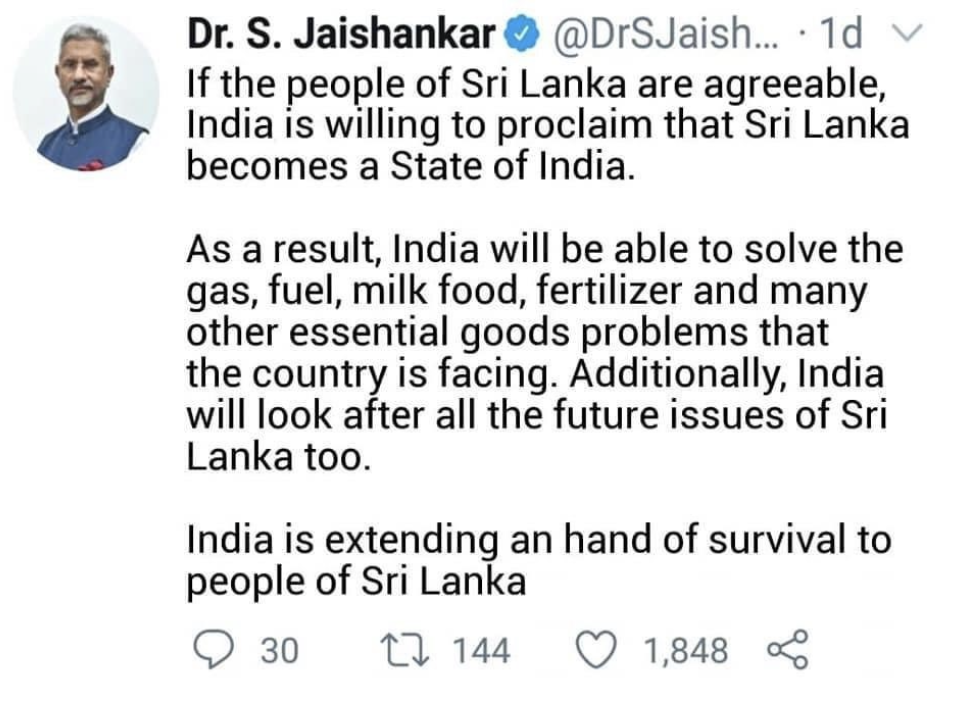}
    \caption{Example of a screenshot where incomplete Twitter handle exists.}
    \label{exception-twitter-handle}
\end{figure}

Similarly, like the timestamp extraction, there are exceptions where it is not possible to extract the complete Twitter handle. For example, Figure \ref{exception-twitter-handle} shows a screenshot example where the Twitter handle is incomplete - `@DrSJaish...'. This incomplete Twitter handle cannot be used to extract the full Twitter handle.

\section{Conclusion}

Screenshots are the easiest way to share content on social media. As there does not exist any specific tool to establish the veracity of  content that is shared as a screenshot, it is a critical task to detect a fabricated post. It is important to verify whether a post is fake or real, because fake posts can be responsible for misinformation and disinformation spread on social media. The collected data set and the tool we are developing for this research would greatly contribute to the research areas of misinformation and disinformation spread on social media.

\begin{acks}
This work supported in part by the GROW M\&S project (Grant \# 300747-010), funded by the US Department of Education.
\end{acks}

\bibliographystyle{ACM-Reference-Format}
\bibliography{sample-base}


\end{document}